# THE EVOLUTIONARY UNIFIED SCHEME.
# I. QUASARS AND RADIO GALAXIES
# IN THE VIEWING ANGLE – REDSHIFT PLANE

F. Vagnetti and R. Spera

Dipartimento di Fisica, Università di Roma "Tor Vergata",
Via della Ricerca Scientifica, I-00133 Roma, Italy
vagnetti@roma2.infn.it, spera@roma2.infn.it

## ABSTRACT

We present a study of the distribution of quasars and radio galaxies in the plane (viewing angle – redshift) in the framework of the *evolutionary unified scheme* (Vagnetti et al. 1991). Results are presented for some illustrative cases, including a distribution of the Lorentz factors, and appropriate luminosity functions for quasars and their host galaxies. A cosmologically increasing $\langle \Gamma \rangle$ is found, in agreement with the previous paper. It is argued that the appearence of sources as quasars or radio galaxies can depend on the viewing angle and on the redshift, due to the balance between the beamed component and the luminosity of the host galaxy. Within the assumptions of our *evolutionary unified scheme*, we find that low-$\Gamma$ objects can be observed as quasars mainly at $z \gtrsim 0.3$, while a substantial fraction of the low-$z$ radio galaxies could consist of quasar-remnants.

*Subject headings*: BL Lacertae objects – cosmology – quasars – radio sources: general

# 1. INTRODUCTION

Strong cosmic evolution is an established property of most classes of Active Galactic Nuclei. A large increase of density and/or luminosity with look-back time is apparent in the luminosity functions of QSOs in the optical (Marshall 1985, Boyle, Shanks & Peterson 1988), as well as in the X-ray bands (Della Ceca et al. 1992, Boyle et al. 1993). A similar behavior is shown by radio loud quasars, which share with radio galaxies strong evolution also in the radio band (Dunlop & Peacock 1990). In fact, evolution was first recognized for radio sources (Longair 1966).

On the other hand, jets and/or other anisotropic phoenomena, often present in the radio sources, have led to the model of the relativistic beam (Blandford & Rees 1978) and later to the so called *unified schemes* (US), which interpret different classes of radio sources as the same objects seen from different directions (see e.g. Browne & Jackson 1992, Antonucci 1993). In fact two separate schemes have been proposed for low-power sources (Browne 1983, Urry & Padovani 1994), and for high-power ones (Barthel 1989, 1994).

But such geometric schemes forget to take into account the evolutionary behavior of the sources. In fact, it has been shown in a previous paper (Vagnetti, Giallongo & Cavaliere 1991, herafter VGC) that the anisotropy of the emission is likely to evolve with the cosmic time, linking in this way geometry and time into an *evolutionary unified scheme* (EUS). Under this hypothesis, the geometric schemes for high- and low-power sources become the starting and ending point of the same evolutionary trend.

The main result of the EUS is that of a slower evolution of the beamed component, compared to the isotropic one, and this effect is attributed to a cosmological (weak) increase of the bulk Lorentz factor of the beam, $\Gamma$. Technical details are briefly recalled in §2.

Our approach to describe interrelations between geometry and evolution is based on the investigation of the (viewing angle – redshift) plane, or $\theta - z$ plane. To study the distribution of sources on such plane, it is necessary to use an appropriate viewing angle probability, as determined by the influence of the relativistic beam (see §3). Some models for the distribution of sources in the $\theta - z$ plane are then presented in §4.

Other evolutionary aspects could be important. For example, the ratio between the evolving quasar luminosity and the luminosity of the host galaxy is of course changing with the cosmic time, and this could influence the identification of the source as a radio loud quasar or a radio galaxy. We will account for this effect in §4.

The intracluster environment of radio sources is also evolving (see e.g. Edge et al. 1990), and this may influence the extended radio structure. It has been argued (see e.g. Browne & Jackson 1992) that an evolutionary connection between high- and low-power radio sources is not likely, because of the different morphological structures, usually referred to as FR II and FR I (Fanaroff & Riley 1974), respectively. In fact, such well-known dichotomy could be related to the different environment of the two classes, FR I sources being generally in richer clusters than FR II (Prestage & Peacock 1988, Yates, Miller & Peacock 1989, Yee & Ellingson 1993).

The physical sizes of radio sources also show cosmological evolution, increasing with cosmic time, and – at a lower degree – with radio power (see e.g. Kapahi 1987, 1989).

Whether the size evolution is similar for radio galaxies and quasars in agreement with US is still a matter of debate (see e.g Gopal-Krishna & Kulkarni 1992, Singal 1993). The size evolution can be understood in terms of change of the environment and of the efficiency with which beam power is converted into radiation (Gopal-Krishna & Wiita 1991). The smaller sizes of high redshift sources suggest a link with the problem of the so called *compact steep-spectrum sources* (CSS) (see e.g. Fanti et al. 1990, Fanti & Fanti 1994), whose position within US should be investigated in an evolutionary way.

Bending of the quasar radio morphologies has also been found to change, showing a decrease with cosmic epoch, possibly caused by interaction of the radio source with an epoch-dependent interstellar or intracluster medium (Barthel & Miley 1988).

Although much progress has already been attained, most of the above properties are likely to depend on the viewing angle, and it is not straightforward that evolution and orientation can be investigated separately. However, only a few aspects will be addressed in the present work. Some points which could be relevant for future investigations are discussed in §5.

## 2. THE EVOLUTIONARY UNIFIED SCHEME

VGC have proposed a connection between flat-spectrum quasars and BL Lac Objects in terms of the *evolutionary unified scheme*. This includes evolutionary change from strong-lined into weak-lined objects, based on the presence of two components of different evolutionary behavior for the optical band:

$$L_{opt}(z) = L_i(z) + L_b(z) \ , \tag{1}$$

i.e an isotropic "thermal" component $L_i$ typical of radio-quiet and steep-spectrum radio-loud quasars, that mainly excites the lines, and a relativistically beamed component[1]

$$L_b = L_j[\Gamma(1 - \beta \cos \theta)]^{-(2+\alpha_{opt})} \ , \tag{2}$$

which dims more slowly to remain dominant at low redshifts, possibly swamping some broad emission lines in BL Lacertae Objects.

In fact, these results derive from a radio-optical analysis on 5 complete samples of quasars with flux greater than 0.1 Jy at 2.7 GHz, including 249 objects (see VGC). Different correlations of the optical-to-radio ratio with radio luminosity and look-back time are found for flat- and steep-spectrum quasars, indicating a slower time-dependence for the former[2]. In turn, such different correlation implies a slower evolution of flat-spectrum quasars in the optical band, possibly due to the effect of the beam. The slower evolution of the beamed component is attributed to a cosmological increase of $\Gamma$, rather than to an evolution of $L_j/L_i$.

VGC then show that there is spectral and statistical continuity between flat-spectrum quasars and BL Lacertae objects, thus suggesting an evolutionary connection between the two classes.

---

[1] Here and in the following we adopt the model of stationary flow in a continuous jet (e.g. Phinney 1985), in which the boosting exponent is $2 + \alpha$ rather than $3 + \alpha$ as in point sources. Spectral indexes defined after $S_\nu \propto \nu^{-\alpha}$.

[2] We must report about a misprint in Table 1 of VGC, not affecting however the results: the numbers in the last column (logarithm of normalization) are incorrect and must be incremented by 1.

In the present paper, we extend the previous study to the whole set of radio sources in the $\theta - z$ plane. Through the investigation of the evolution as a function of the viewing angle, we find again an increasing $\langle \Gamma \rangle$ (cf § 4). We then examine the distribution of quasars and radio galaxies in the $\theta - z$ plane. This will allow (in a distinct paper) the prediction of the distribution of observable quantities, such as superluminal motion and jet asymmetry, separately for quasars and radio galaxies, to test the cosmic increase of $\langle \Gamma \rangle$.

## 3. THE APPARENT DISTRIBUTION OF VIEWING ANGLES

To study the distribution of sources in the $\theta - z$ plane, it is first needed to know how the presence of the beamed component affects the observed distribution of viewing angles. In fact, it has been shown by Cohen (1989, see also Vermeulen & Cohen 1994) that, in a flux-limited sample, low luminosity sources can be observed more easily if their beamed component lies near the line of sight, due to Doppler amplification. The probability of observing sources of small $\theta$ (angle between the beaming axis and the line of sight) is then increased (for given $\Gamma$) according to the following equation:

$$p_{\Gamma}(\theta) \propto (1 - \beta \cos \theta)^{-np} \sin \theta \; , \qquad (3)$$

where $n = 2 + \alpha_r$ is the boosting exponent and $p$ is the slope of the integral counts of the parent population in the radio band, if a radio-selected sample is considered. Eq. 3 holds for the case of a purely beamed source.

For purely isotropic sources the probability is simply proportional to the solid angle: $p(\theta) \propto \sin(\theta)$. More generally, we will consider sources with both beamed and isotropic components, namely, core and extended components. The flux will be given by $S = S_i + S_b = S_i\{1 + f[\Gamma(1 - \beta \cos \theta)]^{-n}\}$, $f$ being the intrinsic fraction of power in the beamed component, and the probability becomes now:

$$p_{\Gamma}(\theta) \propto \{1 + f[\Gamma(1 - \beta \cos \theta)]^{-n}\}^p \sin \theta \; . \qquad (4)$$

The probability has a peak at small $\theta$ due to the beamed component and increases again at large $\theta$ if the isotropic component is not negligible, see fig. 1.

## 4. THE $\theta - z$ PLANE

### 4.1 A Simplified Model

The optical luminosity, as given by eqs. 1 and 2, is assumed in VGC to evolve according to a pure luminosity evolution with exponential form $L_{opt} = L_o e^{kT}$ in the look-back time $T$. Assuming pure exponential luminosity evolution also for $L_i$ and $L_j$, the total luminosity can be rewritten:

$$L_{opt} = L_{io}e^{k_i T} + L_{jo}e^{k_j T}[\Gamma(1 - \beta \cos \theta)]^{-(2+\alpha_{opt})} \; . \qquad (5)$$

Eq. 5, considered as a function of $\theta$, describes the evolution of a population of sources oriented at a given $\theta$; averaged over the appropriate angular intervals, must correspond to the evolution of flat- and steep-spectrum quasars analyzed by VGC:

$$L_{o(f,s)}e^{k_{(f,s)}T} = L_{io}e^{k_i T} + L_{jo}e^{k_j T}\langle[\Gamma(1-\beta\cos\theta)]^{-(2+\alpha_{opt})}\rangle_{(f,s)} \ . \tag{6}$$

Here the subscripts $f$ and $s$ refer to the flat-spectrum and steep-spectrum quasars, respectively. $k_f$ and $k_s$ are taken from VGC. $L_{of}$ and $L_{os}$ correspond to two arbitrary evolutionary tracks for typical flat- and steep-spectrum quasars. The beamed term in eq. 6 has to be averaged, using eq. 4, over the angular intervals corresponding to the two populations. Let us consider the ratio $L_s(z)/L_f(z)$. From eq. 6, with the definitions $c = L_{os}/L_{of}, a = L_{io}/L_{jo}, \Delta k = k_s - k_f, \Delta k_1 = k_i - k_j$, we obtain

$$\frac{L_s(z)}{L_f(z)} = ce^{\Delta kT} = \frac{ae^{\Delta k_1 T} + \langle\delta^{2+\alpha_{opt}}\rangle_s}{ae^{\Delta k_1 T} + \langle\delta^{2+\alpha_{opt}}\rangle_f} \ , \tag{7}$$

$\delta = [\Gamma(1-\beta\cos\theta)]^{-1}$ being the Doppler factor.

If we now assign some values to the various parameters in eq. 7, and if we can evaluate the angular intervals corresponding to the two quasar populations, we can then apply eq. 7 at different redshifts and find $\Gamma(z)$. We assume for the moment, following Barthel (1989), that all quasars are oriented within an angle $\theta_{max} \simeq 45°$. The angle $\theta_{max}$ is determined by anisotropic dust distribution, and assumed to be redshift independent. We further assume that the angular intervals are contiguous and not overlapped, i.e. flat-spectrum quasars lie between 0 and $\theta^*$, and steep-spectrum quasars between $\theta^*$ and $\theta_{max}$. $\theta^*$ can be evaluated counting flat- and steep-spectrum quasars in the same samples already used for the radio-optical analysis (see VGC), so that the integral $\int_0^{\theta^*} p_\Gamma(\theta)d\theta$ gives the observed fraction of flat-spectrum quasars. It is seen that this fraction (considering quasars with $\alpha_r < 0.2$) is nearly constant $\simeq 0.4$ for $z \gtrsim 0.3$, while only flat-spectrum quasars are observed for very low redshifts, see fig. 2. We believe that the striking absence of steep-spectrum quasars at low redshifts is not intrinsic, rather determined by the possible dominance of the luminosity of the host galaxy: due to the presence of the beamed component (eq. 5), it is easier for quasars at large viewing angles to be observed as radio galaxies than for quasars at small angles. It is then possible that many low redshift radio galaxies are in fact steep-spectrum quasars of low luminosity, outshined by their host galaxies. This agrees with the observation of weak nuclear activity in some low redshift radio galaxies (Yee & De Robertis 1989). A similar point of view is also discussed by Yee & Ellingson (1993). We will assume that this is the case and that the "intrinsic" fraction of flat-spectrum quasars is not very different from 0.4 also at $z \lesssim 0.3$:

$$\frac{N_f}{N_Q} = \frac{\int_0^{\theta^*} p_\Gamma(\theta)d\theta}{\int_0^{\theta_{max}} p_\Gamma(\theta)d\theta} \simeq 0.4 \ . \tag{8}$$

We then solve eqs. 7 and 8 together and determine $\Gamma(z)$ and $\theta^*(z)$. Results are presented in fig. 3 for $\Gamma(z)$ corresponding to different values of $\Gamma_o = \Gamma(0)$.

Having determined $\Gamma(z)$, we then rewrite eq. 5 as follows:

$$L(\theta,z) = \frac{ae^{\Delta k_1 T} + [\Gamma(1-\beta\cos\theta)]^{-(2+\alpha_{opt})}}{ae^{\Delta k_1 T} + \langle[\Gamma(1-\beta\cos\theta)]^{-(2+\alpha_{opt})}\rangle_s}L_{so}e^{k_s T} \ . \tag{9}$$

$L(\theta, z)$ is shown in fig. 4. From eq. 9, we can see that low-redshift quasars should be observed as radio galaxies if $L(\theta, z) < L_G$, $L_G$ being the host galaxy luminosity. This possibility depends on both $\theta$ and $z$. The critical angle $\theta_G(z)$ is shown in fig. 5 for an illustrative case. We note that this effect would hold also for $\Gamma = const$, in fact it is based only on two ingredients: (i) the presence of an anisotropic component, and (ii) the evolutionary nature of the quasar luminosity.

## 4.2 Distribution of Lorentz Factors

The next step is to assume a distribution of Lorentz factors. In fact, it has been pointed out by Cohen (1989) that the expected distribution of superluminal velocities for objects with a single $\Gamma$ does not reproduce the observed VLBI data. Impey, Lawrence & Tapia (1991) have then assumed a distribution of the form $n(\Gamma) \propto \Gamma^{-x}$, with $\Gamma_{max} = 10$. Comparison with the data from Cohen gives satisfying fits for $x \lesssim 2$. Padovani & Urry (1992) suggest a distribution in the range $5 < \Gamma < 40$, with $x = 2.3$.

The distribution of viewing angles has to be replaced by the following formula:

$$p^*(\theta) \propto \int_{\Gamma_{min}}^{\Gamma_{max}} p_\Gamma(\theta) \ \Gamma^{-x} d\Gamma \ , \tag{10}$$

which is shown in fig. 6 for some values of the relevant parameters. Eqs. 7 and 8 are then rewritten making use of $p^*(\theta)$. We assume a fixed slope $x = 2$ and allow $\Gamma_{max}$ to vary with $z$, choosing $\Gamma_{max}(0) = 40$ (another possible choice would be to assume a fixed $\Gamma_{max}$ and to obtain a varying slope).

The parameter $f$ can be derived from $R$, the ratio between the core and lobe radio components, assuming that its minimum observed value $R_{min}$ (for a given sample of quasars and radio galaxies) approaches the absolute minimum $R(\theta = 90°, \Gamma = \Gamma_{max}) = 2f\Gamma_{max}^{-(2+\alpha_r)}$ (the factor 2 is due to the contribution of the counter-jet). In fact, due to the limitation of the samples, the observed ratio will be greater then the absolute minimum by a factor which depends on the assumed boosting exponent and $\Gamma$ distribution. For estimating $R_{min}$, we use the compilation by Morganti, Killeen & Tadhunter (1993), where it is found $R_{min} \simeq 10^{-4}$ for a subsample of the 2-Jy sample by Wall & Peacock (1985). It can then be evaluated $f < R_{min}\Gamma_{max}^{2+\alpha_r}/2 \simeq 0.04$. We have computed numerical $R$-distributions, which suggest a decrease of the above value by a factor $\gtrsim 3$ for a sample of $\sim 100$ sources. We finally adopt the value $f = 0.01$. This can be compared with the results of Padovani & Urry (1992), who find $f = 0.007$, from the comparison of luminosity functions of beamed and unbeamed objects.

With such low values of $f$, the viewing angle distribution $p^*(\theta)$ becomes nearly isotropic, so that eq. 8 can be computed with the isotropic distribution $\sin\theta$, giving $\theta^* \sim 28°$ at all redshifts for the angle dividing flat- and steep-spectrum quasars. Eq. 7 must instead be computed with the appropriate distributions $p_\Gamma(\theta)$, eq. 4, which are substantially anisotropic for large $\Gamma$'s. In fact, the terms in brackets of eq. 7 can be written as:

$$\langle \delta^{2+\alpha_{opt}} \rangle = \int d\Gamma \ n(\Gamma) \int d\theta \ p_\Gamma(\theta) \ \delta^{2+\alpha_{opt}} \ , \tag{11}$$

so that the anisotropic modifications due to $p_\Gamma(\theta)$ are greatly enhanced in the product with $\delta^{2+\alpha_{opt}}$.

The critical angle $\theta_G(z)$ which divides quasars from radio galaxies is now different for each $\Gamma$ in the distribution: for example in some intervals of $\theta$ there are objects with large $\Gamma$'s observed as quasars and objects with low $\Gamma$'s observed as radio galaxies. $\theta_G(\Gamma, z)$ is shown in fig. 7 for a typical case. Quasars are observed below the surface and radio galaxies above the surface. We note that low redshift objects with $\Gamma \lesssim 5$ are observed as radio galaxies at all the angles. We further note the abrupt change of $\theta_G$ at $z_{crit} \simeq 0.3$: independently of $\Gamma$, $\theta_G$ attains the value $\theta_{max}$ (we recall that it is assumed $\theta_{max} = 45°$, following Barthel 1989). The effect can be interpreted as follows: at $z > z_{crit}$ the isotropic component is sufficient to outshine the host galaxy, while at lower $z$ only the beamed component can do the job, depending on $\Gamma$ and $\theta$. The effect is so sharp, because in the present calculation we have chosen single values of luminosity for the quasar and the host galaxy. The effect is to be cured adopting appropriate luminosity functions (this is done in the next section). Using $\theta_G(\Gamma, z)$ and the distribution of $\Gamma$, we now compute the expected fraction of quasars in the $\theta - z$ plane, see fig. 8. We note that low redshift quasars (and BL Lac objects[3]) are expected to be seen only at very small angles.

### 4.3 Luminosity Functions of Host Galaxies and Quasars

The discontinuity at $z \sim 0.3$ in figs. 7 and 8 depends on the choice of single luminosities for host galaxies and quasars. The effect occurs when the isotropic component (which can be approximated by the luminosity of steep-spectrum quasars) outshines the luminosity of the host galaxy, therefore when

$$L_s(z) = L_{so} \exp(k_s T) > L_G \; . \tag{12}$$

We therefore need appropriate luminosity functions for host galaxies and for steep-spectrum quasars.

Following the analysis and discussion by Yee (1992) and Yee & Ellingson (1993), we adopt for the host galaxies a luminosity function shaped as a gaussian distribution with $\langle M_B \rangle = -21.8$, $\sigma = 0.45$. This is a representation of the luminosity function of radio galaxies, which – in the spirit of a unified scheme – is shown by Yee to be compatible, within the errors, with the luminosity function of the host galaxies of radio loud quasars.

The luminosity function of steep-spectrum quasars is needed at $z = 0$, with known evolution. As the objects are radio-selected, the optical luminosity function has to be computed through the conditional distribution function and the radio luminosity function (cf. VGC):

$$\Psi_{opt}(L_{opt}) = \int \Psi_R(L_R) \; \Phi \left( \frac{L_{opt}}{L_R} \mid L_R \right) dL_R \; . \tag{13}$$

The result, following again data and analysis of VGC, is a gaussian distribution with $\langle M_B \rangle \simeq -20$, $\sigma = 1.2$.

The expected fraction of quasars in the $\theta - z$ plane is then computed again, convolved through the luminosity functions discussed above. The result is shown in fig. 9. Although the surface is now very smooth, also in this case very few quasars are expected at $z \lesssim 0.3$, except that for small angles.

---

[3] In fact, with the term "quasars" we mean objects in which the nuclear optical component dominates over the host galaxy, and this should include also BL Lac objects.

# 5. DISCUSSION

We have built a framework to investigate the behavior of radio sources in the $\theta - z$ plane. This can be accomplished through the observation of properties which depend strongly on the viewing angle and on the Lorentz factor, such as the superluminal velocity and the asymmetry (flux ratio between the jet and the counterjet). Distributions of asimmetries and of superluminal velocities will be presented in a future paper. We expect a substantial population of low redshift radio galaxies with the properties of hidden quasars and then with high asimmetry and superluminal velocity. Quantitative results will allow a direct comparison of the EUS with the more conventional case of constant $\Gamma$.

Some criticism and/or difficulties should be mentioned. Padovani (1992) has applied the same radio-optical analysis used in VGC to the complete sample of 1 Jy BL Lacs (Stickel et al. 1991) and to a sample of flat-spectrum quasars derived by the 2 Jy sample of Wall & Peacock (1985). A significant correlation, indicating an evolving optical-to-radio ratio, is found by Padovani for BL Lacs, while not for flat-spectrum quasars. It is argued in favor of a possible contamination of BL Lacs in the quasar samples used by us. In fact, we did not even find a significant correlation using the 2 Jy sample alone, due to the very limited range in flux, in agreement with Padovani, but we find it when we use 5 samples down to 0.1 Jy (cf VGC). On the other hand, the significant correlation found for BL Lacs fits very well within EUS, which assumes a continuity between flat-spectrum quasars and BL Lacs. Padovani also reports a comparison of the intrinsic luminosities of [OIII] for *a few* BL Lacs and flat-spectrum quasars: BL Lac objects have intrinsically weaker lines, and it is concluded that their lower equivalent widths are not simply a consequence of a higher continuum. However, we note that the comparison is made in somewhat different ranges of redshift, probably implying some evolution of the intrinsic line luminosity. The main result of EUS is that of an evolution of the optical-to-radio ratio for flat-spectrum quasars, also implying a cosmological increase of $\Gamma$ and of the non-thermal continuum. It is not assured that such increase is the only factor determining decrease of the equivalent width, neither that by itself it is sufficient to outshine the lines.

That BL Lac objects constitute a population physically different from flat-spectrum quasars is asserted by many authors. One interesting argument is reported by Tornikoski et al. (1993), who measure the variability at high radio frequency for a complete sample of radio sources. The (few) BL Lacs in the sample do not show variability timescales as short as quasars do.

Another difficulty is the cosmological increase of the Lorentz factor. It is generally believed that BL Lac objects have lower $\Gamma$'s than quasars, to explain their lower observed proper motions (see e.g. Cohen 1989, Impey 1992). In fact, the surveys for superluminal motions are not complete in any sense. Moreover, the result, if significant, could also derive from a very small viewing angle. Vermeulen & Cohen (1994), using superluminal motion statistics, suggest that $\Gamma$ could be nearly constant, or decreasing with cosmic time, depending on the cosmology; but again the result is based on a dataset that is not homogeneous. Also the subsample of core-selected quasars seems inappropriate to us, in that it doesn't contain galaxies. In fact, sources oriented at the optimal angle for superluminal motion could also be (at low $z$) FR I radio galaxies; this however will be investigated in a future paper. Ghisellini et al. (1993), applying the synchrotron self Compton model, obtain lower Doppler factors for BL Lacs than for core-dominated quasars. Using also information on superluminal motions and on core dominance, they claim that BL Lacs have larger viewing angles, similar $\Gamma$, and higher $f$ than core-dominated quasars, and are therefore a separate population. But: (i) the sample used by Ghisellini et al. is not statistically complete, on the contrary it includes solely the objects, whose radio cores have been measured by VLBI; (ii) Doppler factors derived by the synchrotron self Compton model are lower limits; (iii) the $\delta$-distribution of BL Lacs differs from that of core-dominated quasars for having many objects with $\delta < 1$, buth then $R = f\delta^n < 1$ should result for such objects, so that BL Lacs could not be the core-dominated counterpart of FR I radio galaxies as the authors assume. We argue that the Doppler factors of BL Lacs are probably higher than the values predicted by the synchrotron self Compton model. As for the radio galaxies, Lower $\langle\Gamma\rangle$ is found by Morganti et al. (1994) for FR I galaxies, compared to FR II. However, the analysis is performed on a very limited range of redshift ($z < 0.7$), compared to that of VGC for quasars.

Would the indications of lower Doppler factors for BL Lacs and FR I be strengthened, the radio-optical analysis of VGC should be revised. We believe that the result of an evolving optical-to-radio ratio is rather strong, but some improvement could be attained by: (i) use of core powers instead of total powers, and (ii) exclusion of CSS sources. Unfortunately, such information is not available for the large dataset used by VGC and us. Other possible modifications to the present scheme could be the analysis of more complicated models with more time-dependent parameters: for example, considering evolution of the beamed fraction $f$, or of the quasar visibility angle $\theta_{max}$, could allow different solutions for the evolution of $\Gamma$.

We now clarify and summarize some results of the present work and of the EUS.

1. Evolutionary behavior and anisotropy are both relevant properties of Active Galactic Nuclei, and of radio sources in particular. These properties must be linked together and we have done it through the EUS.

2. The EUS does not contradict the so called high luminosity and low luminosity US, instead it includes them. Also within EUS are FR II radio galaxies the parent population of core-dominated quasars, and FR I the parent population of BL Lacs. There is in addition an evolutionary link between FR II and FR I sources.

3. FR I radio galaxies include – at large viewing angles – obscured sources, remnants of FR II radio galaxies, and – at intermediate viewing angles – host-dominated sources, remnants of steep-spectrum quasars.

4. Canonical BL Lac objects at low redshifts are remnants of flat-spectrum quasars, whith the beam dominating over the host galaxy. Their featureless spectra are explained by a combination of a higher continuum and lower line luminosity. The BL Lacs at intermediate redshift have probably higher $\Gamma$ and therefore higher continuum, compared to the OVV quasars at the same redshifts (cf Gopal-Krishna & Wiita 1993). They could be the objects with the highest $\Gamma$'s within a $\Gamma$ distribution.

5. The presence of the beamed component affects the apparent distribution of viewing angles, as shown by Cohen (1989). We have generalized that result to the case of two components.

6. We predict an increase of the average bulk Lorentz factor $\langle \Gamma \rangle$ with the cosmic time. This point will be investigated in a future paper through detailed distributions of superluminal motion and of jet asymmetry.

7. We have computed the expected quasar fraction in the $\theta - z$ plane. Very few quasars are envisaged at low redshift ($z \lesssim 0.3$), except that at small viewing angles, where mostly BL Lac Objects should be observed.

*Acknowledgments.* We thank P. Padovani, G. C. Perola and D. Trevese for useful discussions. Work performed under grants by MURST and CNR.

## FIGURE CAPTIONS

Fig. 1: Distribution of the viewing angle for a two-component case (single $\Gamma$). The shown distributions are for $\Gamma = 5$ and for $f = 0$ (purely isotropic), $f = 0.3$, and $f = 100$ (which approximates the purely beamed case).

Fig. 2: The cumulative fraction of flat-spectrum quasars as a function of $z$, for the samples analyzed by Vagnetti et al. 1991.

Fig. 3: Evolution of $\Gamma(z)$ for three illustrative models, with $\Gamma_o = 2, 5, 10$, respectively. $f = 0.3, q_o = 0.1$.

Fig. 4: Luminosity evolution for sources oriented at given viewing angles. $\Gamma_o = 5, f = 0.3, q_o = 0.1$. It is seen that the evolution becomes lower for sources at smaller $\theta$. The dotted line corresponds to the luminosity of a typical host galaxy, to show that sources at larger $\theta$ are expected to be outshined by their host galaxies for larger intervals of $z$.

Fig. 5: The $\theta - z$ plane for an illustrative model with $\Gamma_o = 5, f = 0.3, q_o = 0.1$. Continuous curves: $\theta_G(z)$ for two different choices of the galactic luminosity: $M_B = -22.5$ (upper) and $M_B = -21.5$ (lower). Sources should be observed as quasars at $\theta < \theta_G(z)$. The dotted curve separates flat- and steep-spectrum objects.

Fig. 6: Distribution of viewing angles for various cases with $n(\Gamma) \propto \Gamma^{-x}$. $\Gamma_{min} = 1$, $\Gamma_{max} = 40$, $x = 2$.

Fig. 7: The critical angle $\theta_G(\Gamma, z)$ for a typical case. Objects with different $\Gamma$ and $z$ are observed as quasars at angles below the surface, where their nuclear luminosity outshines that of the host galaxy. This is due to the isotropic component at large redshifts (see discussion in the text). The right border of the surface marks $\Gamma_{max}(z)$. For this model $\Gamma_{max}(0) = 40, x = 2, f = 0.01, q_o = 0.1$. The adopted luminosities for host galaxies and quasars are the mean values of the luminosity functions discussed in § 4.3.

Fig. 8: The expected quasar fraction in the $\theta - z$ plane, for the model shown in fig. 7. Low redshift quasars should be seen at very small angles, due to the effect of beams with large $\Gamma$'s. Radio galaxies should be seen at every angle for low $z$, and at $\theta \gtrsim 45°$ at large $z$, following Barthel (1989).

Fig. 9: (a) The expected quasar fraction in the $\theta - z$ plane, after convolution with quasar and galaxy luminosity functions as discussed in § 4.3. (b) The same as (a), shown as isocontours, corresponding to $f_Q = 0.1, 0.2, \ldots 0.7$, from bottom to top.

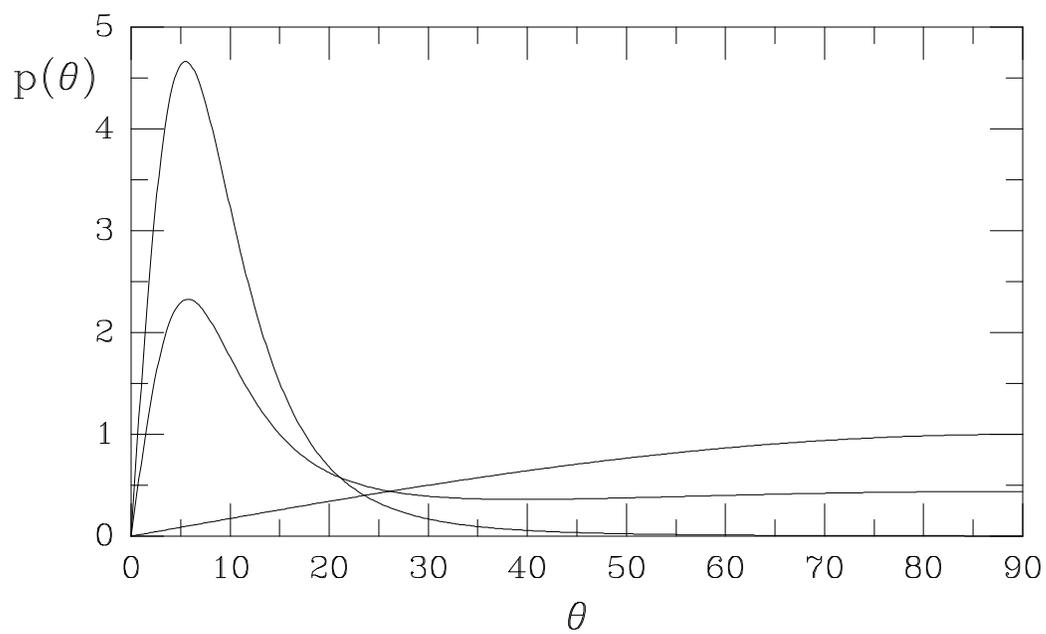

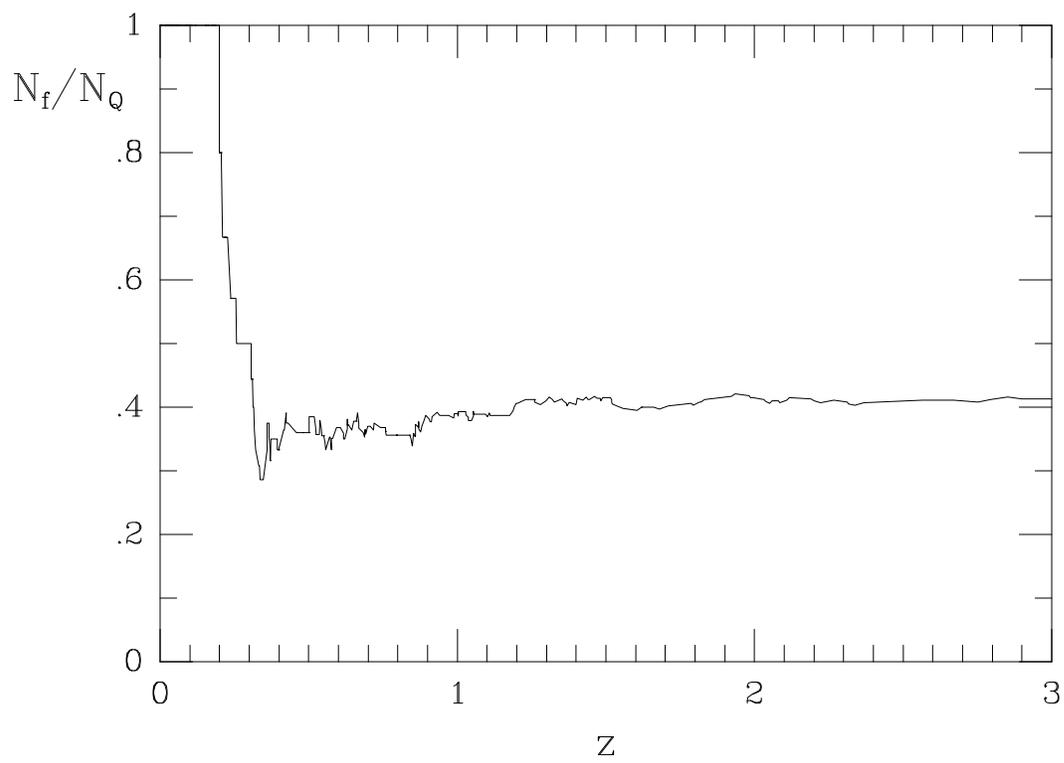

Fig. 1-2

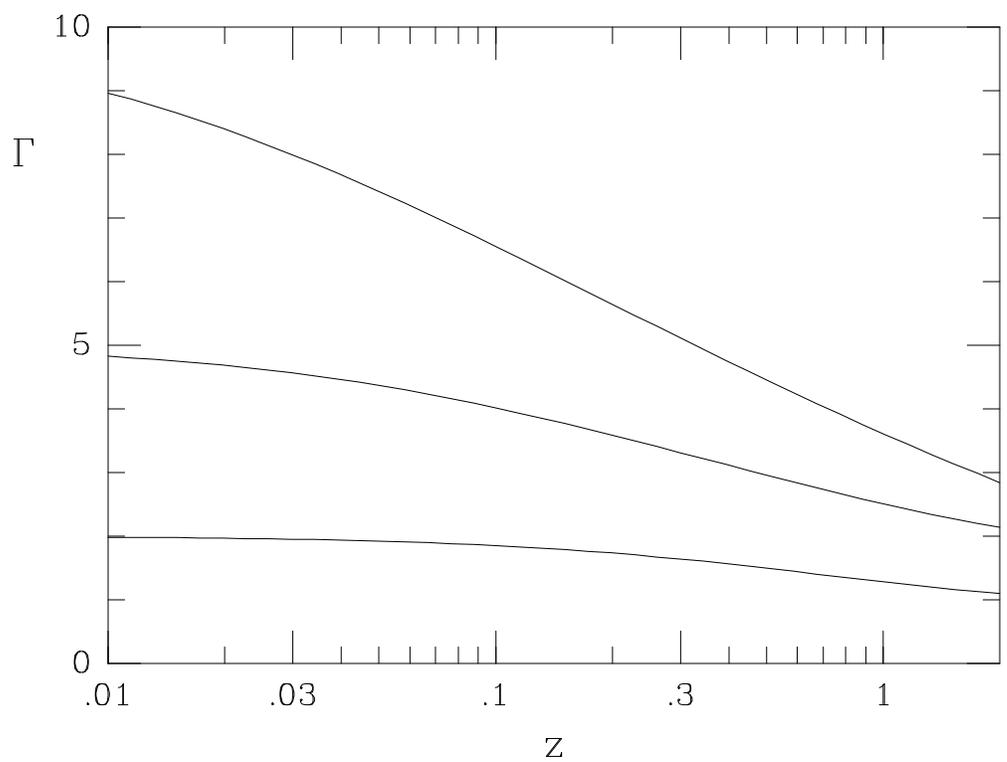

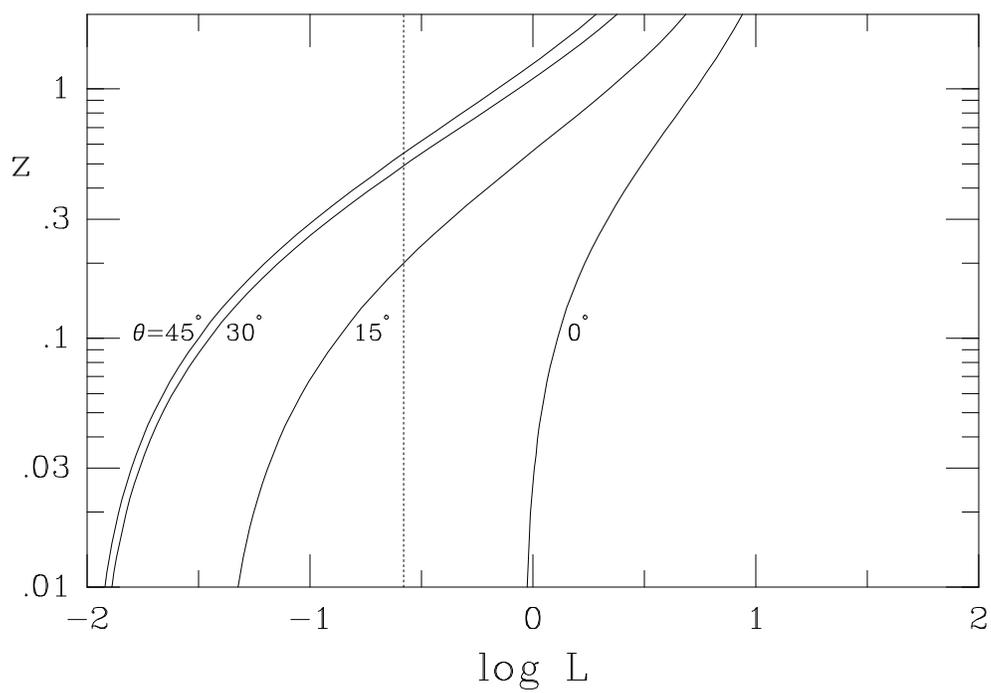

Fig. 3-4

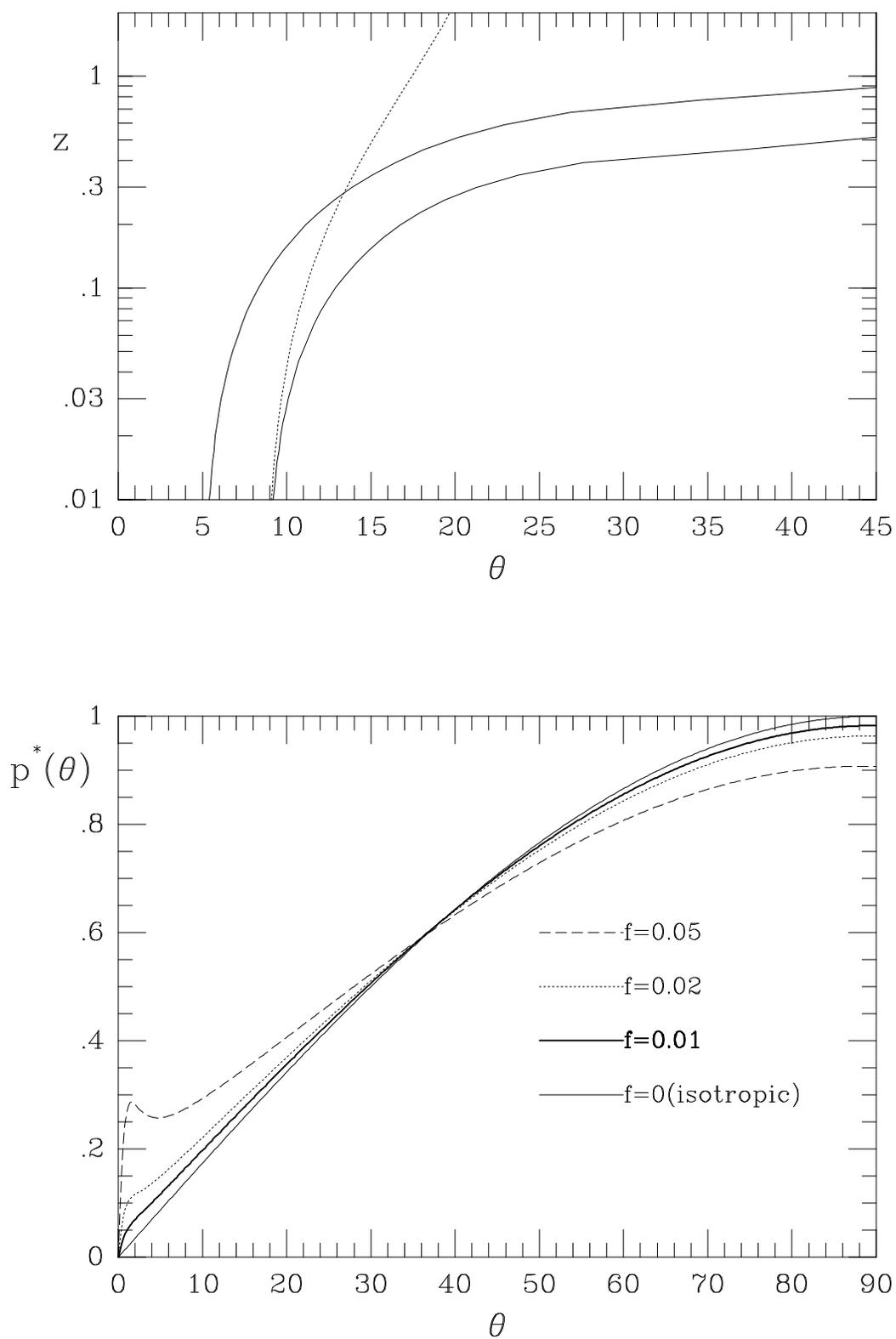

Fig. 5-6

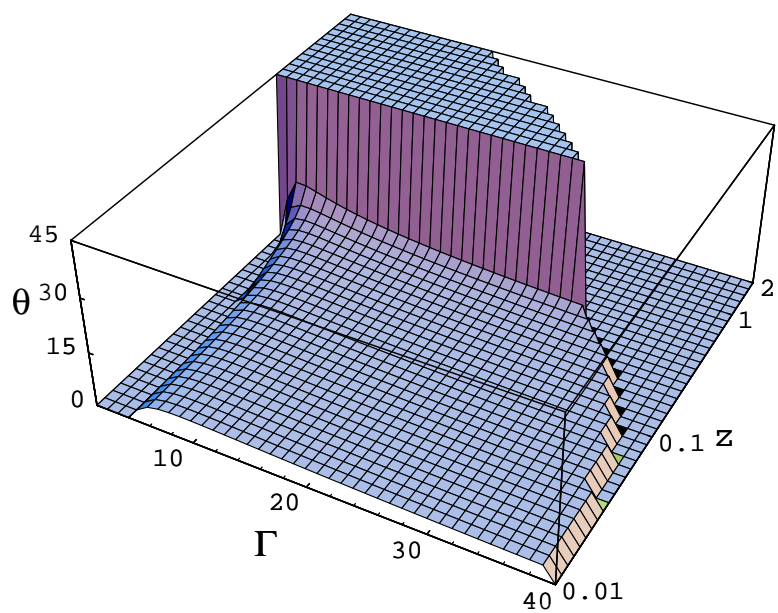

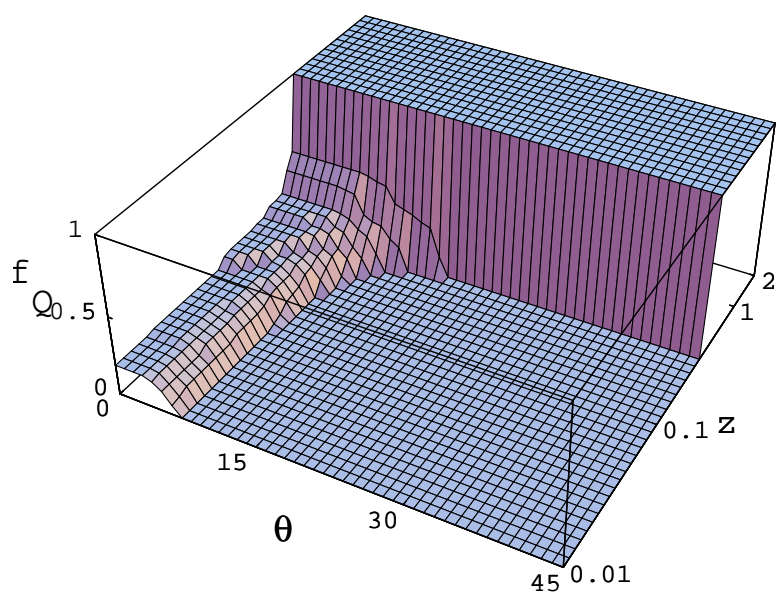

Fig. 7-8

(a)

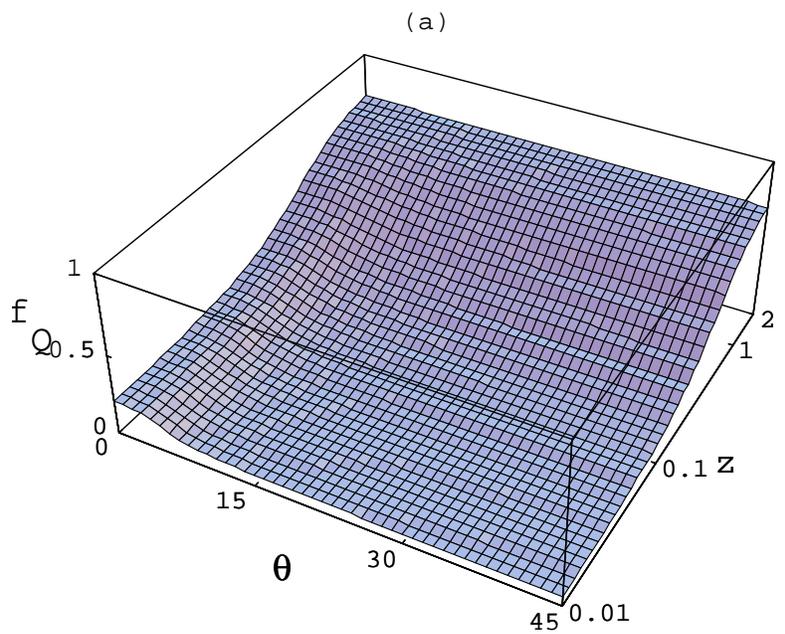

(b)

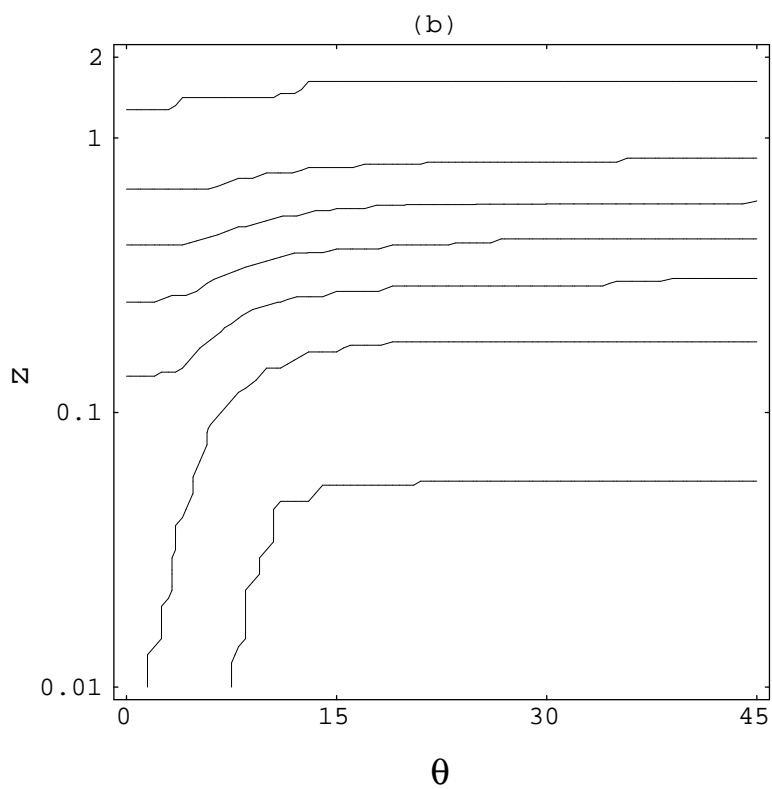

Fig. 9